\documentclass[manuscript,screen]{acmart}
\usepackage{amsmath,amsthm,graphicx,multirow,subcaption}

\makeatletter
\theoremstyle{plain}
\newtheorem{thm}{Theorem}
\newtheorem{defn}{Definition}

\DeclareMathSymbol{:}{\mathord}{operators}{"3A}

\makeatother

\begin{document}
\title{Contextual Integrity Games}

\author[]{Ran Wolff}
\email{ranwolff@amazon.com}

\begin{abstract}
The contextual integrity model is a widely accepted way of analyzing  the plurality of norms that are colloquially called ``privacy norms''. Contextual integrity systematically describes such norms by distinguishing the type of data concerned, the three social agents involved (subject, sender, and recipient) and the transmission principle governing the transfer of information. It allows analyzing privacy norms in terms of their impact on the interaction of those agents with one another. 

This paper places contextual integrity in a strict game theoretic framework. When such description is possible it has three key advantages: Firstly, it allows indisputable utilitarian justification of some privacy norms. Secondly, it better relates privacy to topics which are well understood by stakeholders whose education is predominantly quantitative, such as engineers and economists. Thirdly, it is an absolute necessity when describing ethical constraints to machines such as AI agents. 

In addition to describing games which capture paradigmatic informational norms, the paper also analyzes cases in which the game, per se, does not encourage normative behavior. The paper discusses two main forms of mechanisms which can be applied to the game in such cases, and shows that they reflect accepted privacy regulation and technologies.
\end{abstract}

\maketitle
\section{Introduction}

Contextual Integrity (CI) provides a structured way in which the appropriateness of information transfers can be discussed. CI describes information as flowing from a data subject to a sender and from that sender to a recipient. The flow of information from subject to sender is usually easily justifiable from an ethical perspective because it occurs as a result of their participation in a social context, which is often voluntary and typically benefits both of them. In contrast, the flow from sender to recipient occurs in another social context, to which the subject is often not part, typically does not volunteer for, and potentially does not benefit from.  CI focuses on the  ethics (appropriateness) of this second transfer.

Informational norms regulate the flow of information between sender and recipient. They can be seen as society's way of handling the potential implications of that flow. Every living society inherits a large volume of  informational norms, but also comes up with new ones which correspond to changes in social context. Not surprisingly, the revolutionary speed and scope of information transfer in today's society triggered abundant modifications to informational norms. One of the major changes has been the commercialization of information: Data aggregators have found uses for individual data in domains starting from retail and leading, just recently, to AI training. As evident from the flurry of legislation around the information economy, society is still working out the details of the norms it would like to adopt in the context of commercial information transfers. 

Informational norms can be based off of various ethical frameworks: Privacy has been argued for on the basis of human dignity  \cite{israel_law_review_1992}, as an expression of human autonomy \cite{henkin1974privacy}, or as a conservative response to new media \cite{warrenright,wuest_2021}. In the context of commercialization it makes sense to ground informational norms in a utilitarian framework: If both subject, sender and recipient are exchanging information for their self interest, and if they externalise no cost to the wider society, then the correct informational norm might well be the one which maximizes participants' utility. 

Utilitarianism has long been suggested, and often criticized, as a basis for the discussion of informational norms. Famously, Posner \cite{baker1977posner} used a utilitarian analysis to argue that privacy norms should not be extended. This paper builds on the theoretical criticism of Posner's analysis which has shown \cite{wolff2015emergent} that respect for privacy can emerge as a strategy which leads to a payoff dominant Nash equilibrium in certain games. In other words, that informational-normative behaviors can be based on utilitarianism. The first contribution of this work is the expansion of such privacy-games to three players, corresponding to the subject, sender, and recipient. We present games in which the strategies which lead to the payoff-dominant Nash equilibria are those which are normative according to six different transmission principles: Confidentiality, Mandatory Transfer, Control, Fiduciary Transfer, Notification, and Information Ownership.

The second contribution of this work is the game-theoretic discussion of mechanisms in games where players do not necessarily have a strategy which we would identify as information-normative. We show that accepted social mechanisms can be mapped onto different types of modifications to the game. Specifically we provide two examples of modifications to the payoff: taxation and transfers, and three examples of modification to the information channel: perturbation, bandwidth limitation, and message vetting. Placing all those mechanisms in a common game theoretic framework allows comparing them against one another, which is not always simple without such abstraction.

Considering privacy from a utilitarian ethics perspectives has potential benefits and risks. 
In a voluntary, commercial setting, the benefit of utilitarian analysis are: Firstly, that the participants: engineers, economists, and their managers, are often better educated to understand computational analysis that they are to understand other normative  arguments. Secondly, it permits the encoding of ethical considerations into AI agents, which increasingly replace human judgement in many applications. This is especially important in commercial ethical problems in which some of the key variables, such as the preferences of users, can only be  can only be understood through experimentation and measurement.

The rest of this paper is organized as follows:
Section \ref{sec:definitions} provides the definition of game-theory as we use it here, with special focus on the use of secrets in games, as well as some preliminary games and shorthand which we use throughout the rest of the paper.
Section \ref{sec:games} provides examples of five games in which  informational-normative strategies lead emerge as the Nash equilibrium.
Section \ref{sec:mechanisms} provides examples for mechanisms which drive players to informational-normative strategies. One of those mechanisms complements the Information Ownership transfer principle. 
Section \ref{sec:related} places this work in the context of previous work.
Finally, Section \ref{sec:discussion} discusses some of the interesting questions opened by the mathematical definition of privacy.

\section{\label{sec:definitions}Definitions, Preliminaries, \& Notations}

Consider a game between  three rational players: Alice, who is also
denoted the subject, Bob, the sender, and Carol, the recipient. 
Each player has a set of
choices $A$, $B$, and $C$ and a gain (payoff) function which
depends on the choices of all of the players: $g_{A}:A\times B\times C\rightarrow\mathbb{R}$,
$g_{B}:A\times B\times C\rightarrow\mathbb{R}$, and $g_{C}:A\times B\times C\rightarrow\mathbb{R}$,
respectively. Throughout this paper A, B, \& C are abstract, meaningless, choices: $A=\left\lbrace Top, Middle, Bottom \right\rbrace$, $B=\left\lbrace Near, In-between, Far \right\rbrace$, and $C=\left\lbrace Left, Center, Right \right\rbrace$ which we shorthand to $\left\lbrace T, M, B \right\rbrace$, $\left\lbrace N, I, F \right\rbrace$, and $\left\lbrace L, C, R \right\rbrace$, respectively.

Each player also has a secret, which is a piece of information known
only to that player. Alice's secret is denoted $a$, Bob's $b$, and Carol's  $c$. 
In this paper secrets are single bits which can be 0 or 1 but more informative secrets are equally possible.
The choices of a player can include communicating a secret which that player has to another player. 
Deciding to pass a secret is denoted \emph{sharing} and to withhold the secret is denoted \emph{keeping}. Unless stated otherwise, games have two stages: Players first make all of the decisions which regard to secret sharing and then pick a strategy based on full knowledge of which secrets every player has. In addition to secrets, players in some games can share signals, which are likewise bits whose value is part of the sharing player's choices.

A player's strategy is an algorithm which selects her choices.  
Strategies are encoded as decision trees in c-like code. 
\begin{itemize}
    \item Alice's strategy s of choosing $M$ is denoted $s=M$
    \item Alice's strategy q of choosing $T$ if a secret $a$ is 1 and $B$ if $a$ is 0 is denoted $q=a?T:B$
    \item Alice's strategy r of following strategy $s$ if $b$ is 1 or $q$ if it is zero is denoted $r=b?s:q$ or, equivalently, $r=b?M:a?T:B$
\end{itemize}

Strategies which rely on a secret are indistinguishable from those which rely on random coin flips to any player who does not know that secret.
The model described in this paper only considers deterministic algorithms (pure strategies). Through the use of a sufficient number of secret, the model  can subsume almost any mixed strategy\footnote{The only difference being that probabilities under this model are limited to rational fractions rather than real ones}.
Strategies which rely on a secret are denoted equivalent if other players cannot distinguish between them. E.g., $a?T:B$ and
$a?B:T$ are equivalent for a player who does not have the secret $a$.

A strategy profile is a combination of strategies, one per player,
and is marked by $\left\langle s_{A},s_{B},s_{C}\right\rangle $ where
$s_{A}$ is Alice's strategy, etc. The expected gain from a strategy profile,
$g\left(\left\langle s_{A},s_{B},s_{C}\right\rangle \right)$, is
denoted $\left\langle g_{A},g_{B},g_{C}\right\rangle $ where $g_{A}$
is Alice's gain, etc. and the expectancy is computed on all unknown secrets. A strategy
profile is a Nash equilibrium if all strategy profiles that are different
in but one player's strategy do not increase the gain of that player.
I.e., a strategy profile $\left\langle s_{A},s_{B},s_{C}\right\rangle$ with gain $\left\langle g_{A},g_{B},g_{C}\right\rangle $ is a Nash equilibrium if for any 
strategy profile $\left\langle s'_{A},s_{B},s_{C}\right\rangle$ with gains 
$\left\langle g'_{A},g'_{B},g'_{C}\right\rangle $ it is true that  $g_{A}\geq g'_{A}$,  for any 
strategy profile $\left\langle s_{A},s'_{B},s_{C}\right\rangle$ with gains 
$\left\langle g'_{A},g'_{B},g'_{C}\right\rangle $ it is true that  $g_{B}\geq g'_{B}$,  and for any strategy profile $\left\langle s_{A},s_{B},s'_{C}\right\rangle$ with gains 
$\left\langle g'_{A},g'_{B},g'_{C}\right\rangle $ it is true that  $g_{C}\geq g'_{C}$. A Nash equilibrium is payoff dominant if it provides better or equal gain to all players than all other Nash equilibria in that game.

Given a game, a distributional mechanism is another game in which players have the exact same choices and in which the sum of payoffs for each strategy profile is not larger. Conceptually mechanisms are interventions by an additional player, the government,  who can force transfer of payoff between players and to itself, but cannot provide additional payoff. A communication mechanism, likewise, is a degradation of the secret passing ability of players. In a communication mechanism the government can obstruct message passing but not improve it.

\subsection{Limitations}
The game theoretic model, described above, is one of the simplest which is considered in game theory. As such it is far from an accurate description of human behavior. Two of the more relevant extensions of game theory are for players with limited rationality (able to make mistakes) and players which do not so much know as they believe (e.g., in the value of a secret). We leave developing the theory of more realistic contextual privacy games to further research.

\subsection{Simplified notation}

A binary decomposition of a 3-party game (see,  Sandholm \cite{sandholm2010decompositions}) is a set of 3 simultaneous 2-party games, one for each pair. In each pair, players have the same choices and strategies that they had in the original game. The payoff of every player in the original game is the sum of the payoff in the two games of the decomposition in which that player participates. We refer to those 2-party games as \emph{contexts} and they often indeed correspond to the concept of context in CI.

The typical game described in this paper has three players, every one of which has three choices and at least one secret. The number of strategies by which a player can choose c choices given s binary secrets is $c^{2^s}$. Listing all those strategies is obstructive for readability and impractical in a paper. We choose instead to typically only present the gain for each combination of choices and extend on secret-dependent strategies in the text. When discussing strategy profiles, we typically only analyze one of each set of equivalents. Furthermore, the game is always presented in its decomposed form, in which relations between each pair of players is easier to follow.

\subsection{\label{sub:preliminaries} Preliminaries}

Consider a game between two players: Alice and Carol. Table \ref{tab:Game-of-Secrecy:} depicts the gain of players in all possible strategy profiles. As can be seen in the table, there are five strategy profile which lead to a Nash equilibrium. The first is $\left\langle M,C \right\rangle$ whose gain is $\left\langle 2,2 \right\rangle$ the other are the equivalent strategy profiles $\left\langle a?T:B,c?L:R \right\rangle$, $\left\langle a?T:B,c?R:L \right\rangle$, $\left\langle a?B:T,c?L:R \right\rangle$, and $\left\langle a?B:T,c?R:L \right\rangle$. In each of these equivalent strategies Alice does not know if Carol chooses L or R and therefore computes the expected gain 4. The same is true for Carol who cannot anticipate if Alice chooses T or B.

\begin{table}[htb]
\small
{}%
\begin{tabular}{|c|c|c|c|c|c|c|c|c|c|c|}
\cline{3-11} 
\multicolumn{1}{c}{} &  & \multicolumn{9}{c|}{{Bob}}\tabularnewline
\cline{3-11} 
\multicolumn{1}{c}{} &  & {L} & {C} & {R} & {c?L:C} & {c?C:L} & {c?L:R} & {c?R:L} & {c?C:R} & {c?R:C}\tabularnewline
\hline 
\multirow{9}{*}{{\rotatebox{270}{Alice}}} & {T} & {0,8} & {0,2} & {8,0} & {4,1} & {4,1} & {4,4} & {4,4} & {0,5} & {0,5}\tabularnewline
\cline{2-11} 
 & {M} & {2,0} & \textbf{2,2} & {2,0} & {2,1} & {2,1} & {2,0} & {2,0} & {2,1} & {2,1}\tabularnewline
\cline{2-11} 
 & {B} & {8,0} & {0,2} & {0,8} & {0,5} & {0,5} & {4,4} & {4,4} & {4,1} & {4,1}\tabularnewline
\cline{2-11} 
 & {a?T:M} & {1,4} & {1,2} & {5,0} & {1,3} & {1,3} & {3,2} & {3,2} & {3,1} & {3,1}\tabularnewline
\cline{2-11} 
 & {a?M:T} & {1,4} & {1,2} & {5,0} & {1,3} & {1,3} & {3,2} & {3,2} & {3,1} & {3,1}\tabularnewline
\cline{2-11} 
 & {a?T:B} & {4,4} & {0,2} & {4,4} & {2,3} & {2,3} & \textbf{4,4} & \textbf{4,4} & {2,3} & {2,3}\tabularnewline
\cline{2-11} 
 & {a?B:T} & {4,4} & {0,2} & {4,4} & {2,3} & {2,3} & \textbf{4,4} & \textbf{4,4} & {2,3} & {2,3}\tabularnewline
\cline{2-11} 
 & {a?M:B} & {5,0} & {1,2} & {1,4} & {3,1} & {3,1} & {3,2} & {3,2} & {1,3} & {1,3}\tabularnewline
\cline{2-11} 
 & {a?B:M} & {5,0} & {1,2} & {1,4} & {3,1} & {3,1} & {3,2} & {3,2} & {1,3} & {1,3}\tabularnewline
\hline 
\end{tabular}

\caption{Game of Privacy: Both players have a strategy of keeping their secret and respecting the other player's privacy. With secrets, both $\left\langle M,C\right\rangle$ and $\left\langle a?T:B,c?L:R\right\rangle$ (and its equivalents) are Nash equilibria. If Alice shares her secret with Carol then Carol can choose $a?L:R$ when Alice chooses $a?T:B$. This leaves $\left\langle M,C\right\rangle$ to be the only Nash equilibrium.}
\label{tab:Game-of-Secrecy:}
\end{table}

We first use this game to exemplify the strategies of secrecy and of respect to privacy.
\begin{defn}
\label{def:secrecy}{[}Secrecy (adapted from \cite{dighe2009secrecy}){]} A player has a strategy of secrecy in a game if she chooses not to share her secret.
\end{defn}
\begin{thm}
    \label{thm:secrecy} If Alice has the choice to keep or share her secret with Carol then the strategy in which Alice keeps her secret is dominant.
\end{thm}
\begin{proof}
    If Alice shares her secret with Carol then she can use it in his strategy. The strategy profile $\left\langle a?T:B,c?L:R \right\rangle$ is no longer a Nash equilibrium because Carol can replace her strategy to $a?L:R$ and increase her gain from 4 to 8. The same is true for the equivalent strategy profiles. The only remaining Nash equilibrium is $\left\langle M,C \right\rangle$, which is less gainful for Alice. Keeping her secret is a payoff dominant strategy for Alice.
\end{proof}

\begin{defn}
    \label{def:respect}{[}Respect for privacy (adapted from \cite{wolff2015emergent}){]} A player has a strategy of respect for privacy if she chooses not to observe another player's secret.
\end{defn}
\begin{thm}
    \label{thm:respect} If Carol has the choice if she observes Alice's secret or respects her privacy then the strategy in which Carol respects Alice's privacy is dominant.
\end{thm}
\begin{proof}
    The proof of Thm. \ref{thm:secrecy} holds regardless if Alice shares his secret willingly or if Carol observes Alice's secret without her consent. 
\end{proof}


\section{\label{sec:games}Informational normative strategies}

With the definitions of games and secrets above, we can now describe a set of games which model informational normative behavior. 
Each of the following sections deals with a different transmission principle: Confidentiality, Mandated Transfer, Control, Notification, and Fiduciary Transfer. Information Ownership is discussed in Section \ref{sec:mechanisms}. 
For each of the transmission principles we describe a real-life example, a game theoretic definition of the normative behavior, and an example of a game in which the normative strategy leads to the payoff dominant Nash equilibrium. 

\subsection{Confidentiality}

\begin{table}[htb]
\small
    \begin{subtable}[h]{0.36\textwidth}
    \centering
    \begin{tabular}{|c|c|c|c|c|}
        \cline{3-5} 
        \multicolumn{1}{c}{} &  & \multicolumn{3}{c|}{{Carol}}\tabularnewline
        \cline{3-5} 
        \multicolumn{1}{c}{} &  & {L} & {C} & {R}\tabularnewline
        \hline 
        \multirow{3}{*}{{\rotatebox{270}{Alice}}} 
           & {T} & {0,16} & {0,4} & {16,0}\tabularnewline
        \cline{2-5} 
            & {M} & {4,0} & {4,4} & {4,0}\tabularnewline
        \cline{2-5} 
        & {B} & {16,0} & {0,4} & {0,16}\tabularnewline
        \hline 
    \end{tabular}
    \caption{Alice v. Carol}
    \label{tab:Confidentiality-Alice-v.-Carol}
    \end{subtable}
    \hfill
    \begin{subtable}[h]{0.27\textwidth}
    \centering
    \begin{tabular}{|c|c|c|c|c|}
        \cline{3-5}
        \multicolumn{1}{c}{} &  & \multicolumn{3}{c|}{{Bob}}\tabularnewline
        \cline{3-5}
        \multicolumn{1}{c}{} &  & {N} & {I} & {F}\tabularnewline
        \hline 
        \multirow{3}{*}{{\rotatebox{270}{Alice}}} 
        & {T} & {2,2} & {0,0} & {0,0}\tabularnewline
        \cline{2-5}
            & {M} & {0,0} & {2,2} & {0,0}\tabularnewline
        \cline{2-5}
            & {B} & {0,0} & {0,0} & {2,2}\tabularnewline
        \hline 
    \end{tabular}
    \caption{Alice v. Bob}
    \label{tab:Confidentiality-Alice-v.-Bob}
    \end{subtable}
    \hfill
    \begin{subtable}[h]{0.27\textwidth}
    \centering
    \begin{tabular}{|c|c|c|c|c|}
        \cline{3-5}
        \multicolumn{1}{c}{} &  & \multicolumn{3}{c|}{{Carol}}\tabularnewline
        \cline{3-5}
        \multicolumn{1}{c}{} &  & {L} & {C} & {R}\tabularnewline
        \hline 
        \multirow{3}{*}{{\rotatebox{270}{Bob}}} 
         & {N} & {0,8} & {0,2} & {8,0}\tabularnewline
        \cline{2-5}
            & {I} & {2,0} & {2,2} & {2,0}\tabularnewline
        \cline{2-5}
            & {F} & {8,0} & {0,2} & {0,8}\tabularnewline
        \hline 
    \end{tabular}
    \caption{Bob v. Carol}
    \label{tab:Confidentiality-Bob-v.-Carol}
    \end{subtable}
\caption{Game of Confidentiality}
\label{tab:Game-of-Confidentiality}
\end{table}

Confidentiality is called for in real life when Alice (the subject)
has information which could help her interact with Bob (the sender) but which might harm her if it becomes known to Carol (the recipient). 
In such scenarios, Alice would be keen to share the secret with Bob if she can trust that revealing
the secret to Carol is somehow harmful to Bob as well. 

In terms of game theory we define a confidential strategy as follows:
\begin{defn}
\label{def:confidentiality}{[}Confidentiality{]} Bob has a strategy of confidentiality towards Alice if whenever she shares her secret a with Bob his strategy is to keep a from Carol.
\end{defn}

To see that confidentiality is an emergent property in some games, consider the Game of Confidentiality in Table \ref{tab:Game-of-Confidentiality}.  The context of Alice and Carol, Table \ref{tab:Confidentiality-Alice-v.-Carol}, is a privacy game, similar to that in Table \ref{tab:Game-of-Secrecy:}. The same is true for the context of Bob and Carol. Thus, in those two games players maximize their payoff if they each make choices the other player cannot anticipate. So when each player has their own secret it is straightforward that the payoff-dominant Nash equilibrium of the Game of Confidentiality is $\left\langle a?T:B, b?N:F, c?L:R\right\rangle$ and the gain in that Nash equilibrium is $\left\langle 9,5,12\right\rangle$.

In $\left\langle a?T:B, b?N:F, c?L:R\right\rangle$ , Alice and Bob each gain 1 from their context. If Alice decides to share her secret then Bob can replace his strategy with one that relies on $a$ rather than on $b$ . Because Carol can no more anticipate $a$ than she can anticipate $b$  the change does not affect Bob and Carol's context. Hence, $\left\langle a?T:B, a?N:F, c?L:R\right\rangle$, in which Alice and Bob are paid 2 each in their context and the total payoff is  $\left\langle 10,6,12\right\rangle$, is a payoff-dominant Nash equilibrium when Alice shares her secret with Bob.

Last, if Bob chooses to share Alice's secret with Carol then Alice would no longer choose a strategy which relies on that secret. If she does, then Carol will always match her choice to Alice's and gain 16 in their context while Alice gains 0 in that context. A payoff of  16  dominates any other choice Carol can make in her other context, and the certainty of gaining 0 makes this strategy inferior to choosing M no matter what payoff Alice earns in her context with Bob. Hence Alice would choose M when Carol has her secret. When Alice chooses M, Carol can still use her own secret, $c$ , to choose between L and R, or she can choose C. In the first strategy Bob would choose N or F using $b$ and in the other he would choose I. 

\begin{thm}
  In the Game of Confidentiality Bob has a strategy of Confidentiality towards Alice
\end{thm}
\begin{proof}
    As we have seen, if Alice shares her secret with Bob then the payoff-dominant Nash equilibrium is $\left\langle a?T:B, a?N:F, c?L:R\right\rangle$, in which Bob's payoff is 6. If Bob shares Alice's secret with Carol then the two possible Nash equilibria, $\left\langle M,b?N:F,c?L:R\right\rangle$ and $\left\langle M,I,C\right\rangle$ both pay Bob 4. Therefore, Bob would not share Alice's secret with Carol. 
\end{proof}

\subsection{Mandatory transfer}

\begin{table}[htb]
\small
    \begin{subtable}[h]{0.3\textwidth}
    \centering
    \begin{tabular}{|c|c|c|c|c|}
        \cline{3-5}
        \multicolumn{1}{c}{} &  & \multicolumn{3}{c|}{{Carol}}\tabularnewline
        \cline{3-5}
        \multicolumn{1}{c}{} &  & {L} & {C} & {R}\tabularnewline
        \hline 
        \multirow{3}{*}{{\rotatebox{270}{Alice}}} 
            & {T} & {0,6} & {0,2} & {10,0}\tabularnewline
        \cline{2-5}
            & {M} & {2,0} & {2,2} & {2,0}\tabularnewline
        \cline{2-5}
            & {B} & {10,0} & {0,2} & {0,6}\tabularnewline
        \hline 
    \end{tabular}
    \caption{Alice v. Carol}
    \label{tab:mandated-Alice-v.-Carol}
    \end{subtable}
    \hfill
    \begin{subtable}[h]{0.3\textwidth}
    \centering
    \begin{tabular}{|c|c|c|c|c|}
        \cline{3-5}
        \multicolumn{1}{c}{} &  & \multicolumn{3}{c|}{{Bob}}\tabularnewline
        \cline{3-5}
        \multicolumn{1}{c}{} &  & {N} & {I} & {F}\tabularnewline
        \hline 
        \multirow{3}{*}{{\rotatebox{270}{Alice}}} 
        & {T} & {0,0} & {0,0} & {0,0}\tabularnewline
        \cline{2-5}
            & {M} & {0,0} & {1,1} & {0,0}\tabularnewline
        \cline{2-5}
            & {B} & {0,0} & {0,0} & {0,0}\tabularnewline
        \hline 
    \end{tabular}
    \caption{Alice v. Bob}
    \label{tab:mandated-Alice-v.-Bob}
    \end{subtable}
    \hfill
    \begin{subtable}[h]{0.3\textwidth}
    \centering
    \begin{tabular}{|c|c|c|c|c|}
        \cline{3-5}
        \multicolumn{1}{c}{} &  & \multicolumn{3}{c|}{{Carol}}\tabularnewline
        \cline{3-5}
        \multicolumn{1}{c}{} &  & {L} & {C} & {R}\tabularnewline
        \hline 
        \multirow{3}{*}{{\rotatebox{270}{Bob}}}
        & {N} & {0,0} & {0,0} & {0,0}\tabularnewline
        \cline{2-5}
            & {I} & {0,0} & {1,1} & {0,0}\tabularnewline
        \cline{2-5}
            & {F} & {0,0} & {0,0} & {0,0}\tabularnewline
        \hline 
    \end{tabular}
    \caption{Bob v. Carol}
    \label{tab:mandated-Bob-v.-Carol}
    \end{subtable}
\caption{Game of Mandatory Transfer}
\label{tab:Game-of-Mandated}
\end{table}

A common real-life example of mandatory transfer is reporting of illicit conduct. For instance, judges in the US (senders) are ethically required  to inform federal and state authorities (recipient) if they find that a witness or a party (subject) evades tax. Unlike confidentiality, we cannot assume that the subject willingly shares the secret of tax evasion with the sender and the question of whether revealing that secret in court is ethically justified is an interesting one. Still, once the judge knows the secret, it is normative (although hardly guaranteed \cite{dollinger2014judicial}) that she will transfer it to the tax authorities. 

\begin{defn}
\label{def:mandated}{[}Mandatory transfer{]} Bob has a strategy of mandatory transfer towards Alice if whenever she shares  her secret a with Bob his strategy is to share a with Carol.
\end{defn}

Consider the Game of Mandatory Transfer in Table \ref{tab:Game-of-Mandated}.  The context of Alice and Carol (Table \ref{tab:mandated-Alice-v.-Carol}) is a privacy game similar to that in Table \ref{tab:Game-of-Secrecy:} in which the expected payoff for Alice is 5 and for Carol is 3, so long as they each choose  using their secret. Bob's  only strategy which pays him more than 0 in that case is I, and the payoff dominant Nash equilibrium is $\left\langle a?T:B,I,c?L:R\right\rangle$ with a payoff of $\left\langle 5,1,3\right\rangle$. 
If Carol somehow  gains access to Alice's secret then the only remaining Nash equilibrium remaining is $\left\langle M,I,C \right\rangle$ whose payoff is $\left\langle 3,3,3\right\rangle$. 

\begin{thm}
    In the Game of Mandatory Transfer Bob has a strategy of Mandatory Transfer towards Alice.
\end{thm}
\begin{proof}
    If Bob somehow learns Alice's secret then his options are to keep it from Carol and continue to receive a payoff of 1 or share the secret with Carol and receive a payoff of 3.
\end{proof}

\subsection{\label{sec:fiduciary}Fiduciary Transfer}

\begin{table}[htb]
\small
  \begin{subtable}[h]{0.24\textwidth}
    \centering
    \begin{tabular}{|c|c|c|c|c|}
        \cline{3-5}
        \multicolumn{1}{c}{} &  & \multicolumn{3}{c|}{{Carol}}\tabularnewline
        \cline{3-5}
        \multicolumn{1}{c}{} &  & {L} & {C} & {R}\tabularnewline
        \hline 
        \multirow{3}{*}{{\rotatebox{270}{Alice}}} 
            & {T} & {-8,8} & {0,2} & {8,-8}\tabularnewline
        \cline{2-5}
            & {M} & {2,0} & {2,2} & {2,0}\tabularnewline
        \cline{2-5}
            & {B} & {8,-8} & {0,2} & {-8,8}\tabularnewline
        \hline 
    \end{tabular}
    \caption{Alice v. Carol -- competitive}
    \label{tab:Fiduciary-Alice-v.-Carol-1}
    \end{subtable}
    \hfill
    \begin{subtable}[h]{0.24\textwidth}
    \centering
    \begin{tabular}{|c|c|c|c|c|}
        \cline{3-5}
        \multicolumn{1}{c}{} &  & \multicolumn{3}{c|}{{Carol}}\tabularnewline
        \cline{3-5}
        \multicolumn{1}{c}{} &  & {L} & {C} & {R}\tabularnewline
        \hline 
        \multirow{3}{*}{{\rotatebox{270}{Alice}}} 
            & {T} & {8,8} & {0,2} & {-8,-8}\tabularnewline
        \cline{2-5}
            & {M} & {2,0} & {2,2} & {2,0}\tabularnewline
        \cline{2-5}
            & {B} & {-8,-8} & {0,2} & {8,8}\tabularnewline
        \hline 
    \end{tabular}
    \caption{Alice v. Carol -- collaborative}
    \label{tab:Fiduciary-Alice-v.-Carol-2}
    \end{subtable}
    \hfill
    \begin{subtable}[h]{0.24\textwidth}
    \centering
    \begin{tabular}{|c|c|c|c|c|}
        \cline{3-5}
        \multicolumn{1}{c}{} &  & \multicolumn{3}{c|}{{Bob}}\tabularnewline
        \cline{3-5}
        \multicolumn{1}{c}{} &  & {N} & {I} & {F}\tabularnewline
        \hline 
        \multirow{3}{*}{{\rotatebox{270}{Alice}}} 
            & {T} & {5,3} & {0,0} & {4,2}\tabularnewline
        \cline{2-5}
            & {M} & {0,0} & {0,0} & {0,0}\tabularnewline
        \cline{2-5}
            & {B} & {4,2} & {0,0} & {5,3}\tabularnewline
        \hline 
    \end{tabular}
    \caption{Alice v. Bob}
    \label{tab:Fiduciary-Alice-v.-Bob}
    \end{subtable}
    \hfill
    \begin{subtable}[h]{0.26\textwidth}
    \centering
    \begin{tabular}{|c|c|c|c|c|}
        \cline{3-5}
        \multicolumn{1}{c}{} &  & \multicolumn{3}{c|}{{Carol}}\tabularnewline
        \cline{3-5}
        \multicolumn{1}{c}{} &  & {L} & {C} & {R}\tabularnewline
        \hline 
        \multirow{3}{*}{{\rotatebox{270}{Bob}}} 
        & {N} & {-15,17} & {0,0} & {17,-15}\tabularnewline
        \cline{2-5}
            & {I} & {0,0} & {0,0} & {0,0}\tabularnewline
        \cline{2-5}
            & {F} & {17,-15} & {0,0} & {-15,17}\tabularnewline
        \hline 
    \end{tabular}
    \caption{Bob v. Carol}
    \label{tab:Fiduciary-Bob-v.-Carol}
    \end{subtable}
    
\caption{Game of Fiduciary Transfer}
\label{tab:Game-of-Fiduciary}
\end{table}

\begin{table}[htb]
\small
    \centering
    \begin{tabular}{|c|c|c|c|c|}
        \hline
         & Shared secrets & Alice \& Carol & Nash equilibrium & Payoffs \tabularnewline
        \hline 
            \multirow{ 2}{*}{I}   & \multirow{ 2}{*}{None} & Competitive & $\left\langle a?T:B,b?N:F,C\right\rangle$ & $\left\langle 4.5,2.5,2\right\rangle$ \tabularnewline
        \cline{3-5}
             &  & Collaborative & $\left\langle a?T:B,b?N:F,C\right\rangle$ & $\left\langle 4.5,2.5,2\right\rangle$ \tabularnewline
        \hline
           \multirow{ 2}{*}{II} & \multirow{ 2}{*}{$a \rightarrow Bob$} & Competitive & $\left\langle a?T:B,a?N:F,c?L:R\right\rangle$ &  $\left\langle 5,3,2 \right\rangle$ \tabularnewline
      \cline{3-5}
             &  & Collaborative & $\left\langle a?T:B,a?N:F,c?L:R\right\rangle$ &  $\left\langle 5,3,2 \right\rangle$ \tabularnewline
        \hline
            \multirow{ 2}{*}{III}  & \multirow{ 2}{*}{$a \rightarrow Bob,Carol$} & Competitive & $\left\langle M,I,C\right\rangle$ &  $\left\langle 2,0,2 \right\rangle$ \tabularnewline
       \cline{3-5}
             &  & Collaborative & $\left\langle M,I,C\right\rangle$ &  $\left\langle 2,0,2 \right\rangle$ \tabularnewline
        \hline
            \multirow{ 2}{*}{IV}  & $a \rightarrow Bob,Carol$  & Competitive & $\left\langle M,I,C\right\rangle$ &  $\left\langle 2,0,2 \right\rangle$ \tabularnewline
        \cline{3-5}
            &    $c \rightarrow Alice$ & Collaborative & $\left\langle c?T:B,b?N:F,c?L:R\right\rangle$ &  $\left\langle 12.5,3.5,9 \right\rangle$ \tabularnewline
        \hline
    \end{tabular}
    \caption{Payoff-dominant Nash equilibria}
    \label{tab:Fiduciary-payoff-dominant}
\end{table}

Real life examples of fiduciary transfer are those in which the subject relies on a better informed sender to manage her secret for her: Professional advisors, lawyers, etc. are often the sender in those examples. Those advisors have some information about the context which the subject does not have. E.g., a lawyer knows not just the law, but also the relevant precedents. In a normative setting, a subject should be able to reveal her secret to the advisor knowing that the advisor will only share that secret if it benefits the subject. In other words, normative fiduciary transfer happens when the sender is better informed than the subject and their interests align.

\begin{defn}
\label{def:fiduciary}{[}Fiduciary Transfer{]} Bob has a strategy of fiduciary transfer towards Alice if whenever she shares her secret a with Bob his strategy is to share a with Carol if and only if sharing will increase Alice's gain.
\end{defn}

In the game of Fiduciary transfer, Table \ref{tab:Game-of-Fiduciary}, Alice and Carol do not know if their context  is competitive (Table \ref{tab:Fiduciary-Alice-v.-Carol-1}) or collaborative (Table \ref{tab:Fiduciary-Alice-v.-Carol-2}). If collaborative, then Alice and Carol's gain can increase if they share a secret whereas in a competitive context it will not. Alice makes her choice about secret sharing without knowing the context. Unlike Alice and Carol, Bob does know which context Alice and Carol have. When choosing if he shares Alice's secret with Carol he can consider what Alice and Carol would do after they learn their context. After all of the players executed their secret sharing strategy they all learn the context and choose their best strategy considering the secrets they know.

The key feature of the Game of Fiduciary Transfer is that Bob will not choose N or F in a way that Carol can predict, and neither will Carol choose L or R in a way that Bob can predict. If Bob chooses N, for example, then Carol's dominant strategy would always be to choose L. That combination is so harmful to Bob that no other payoff he may get from Alice can make it preferable to choosing I. Hence, for Alice and Carol to collaborate, if at all their context is collaborative, Carol must share her secret with Alice.

Table \ref{tab:Fiduciary-payoff-dominant} presents the payoff-dominant Nash equilibrium in four different secret sharing scenarios: Firstly, if players each have their own secret then, regardless of Alice and Carol's context, their payoff-optimal Nash equivalence is $\left\langle a?T:B,b?N:F,c?L:R\right\rangle$ with a payoff of $\left\langle 4.5, 2.5,2\right\rangle$. If Alice shares her secret with Bob then they can coordinate by choosing $\left\langle a?T:B,a?N:F,c?L:R\right\rangle$. That would increase their payoffs to  $\left\langle 5, 3,2\right\rangle$ without altering their position in their respective contexts with Carol, who does not know $a$. 
 
 When Bob knows Alice's secret he can choose if he shares it with Carol. By sharing Alice's secret with Carol,  Bob determines that Alice would never be able to make a choice that Carol cannot anticipate. If Alice and Carol are in a competitive context then that forces Alice to choose M because her potential gain from Bob cannot compete with her certified loss to Carol if she makes another choice. If Alice chooses M, then Carol would prefer to choose C and gain 2 in their context over choosing L or R based on her secret and gaining 1 from Bob. Hence,  $\left\langle M, I,C\right\rangle$, with a payoff of  $\left\langle 2, 0,2\right\rangle$,  is the payoff dominant Nash equilibrium.

 Last, if Bob shares Alice's secret with Carol and if Alice and Carol are in a collaborative context then Alice and Carol can collaborate by choosing T and L or B and R. However, as we have seen Carol cannot collaborate with Alice in a way which Bob can predict. So Carol cannot just choose L (or R) and cannot rely on Alice's secret, which is known to Bob. Carol's only way out is to share her own secret with Alice. In this way,  Alice and Carol can both use Carol's secret which Bob does not know. If they do then Bob's best strategy is to choose N or F in a way Carol cannot anticipate. Therefore  $\left\langle c?T:B,b?N:F,c?L:R\right\rangle$ is the payoff-dominant Nash equilibrium, with gain  $\left\langle 12.5, 3.5,9\right\rangle$.
 \begin{thm}
     In the Game of Fiduciary Transfer Bob has a strategy of fiduciary transfer towards Alice.
 \end{thm}
 \begin{proof}
     If Alice and Carol's context is competitive then by sharing Alice's secret to Carol Bob reduces Alice's payoff from 5 to 2 and his own revenue from 3 to 0. If the context is collaborative then by sharing that secret Bob causes Carol to share her own secret with Alice. This increases Alice's payoff from 5 to 12.5 and Bob's payoff from 3 to 3.5. Therefore, Bob would only share Alice's secret with Carol when that benefits Alice.
 \end{proof}

\subsection{\label{subsec:Control}Control}

An example of a control norm in a commercial setting are cases in which a customer must communicate with a service provider through the service of another company.
For instance, the user of a Website who communicates with an advertiser through the publisher Website.
The user may choose to share some information with the publisher (e.g., their delivery address) in order to gain some of their services. 
The publisher often does not know if sharing that information with the advertiser would benefit or harm the user. It is normative for the publisher to ask users if they can share their information and to follow their dictates.

\begin{defn}
\label{def:control}{[}Control{]} Bob has a strategy of control towards Alice if when she shares her secret a and signal s with Bob his strategy is to share a with Carol if s=1 and keep it if s=0.
\end{defn}

Signalling only makes sense if Alice has information which Bob does not have.  Suppose that with the same setup as in the Game of Fiduciary Transfer (Table \ref{tab:Game-of-Fiduciary}) it is now Alice who knows her context with Carol and Bob who does not know.

Assume Alice can choose a signal as  part of her secret sharing strategy.  When she decides to share $a$ she also shares $s$ which she chooses to be 0 or 1. Bob's secret sharing strategy can rely on $s$. So when Alice shares her secret with Bob his choices are: 1) To always keep $a$ from Carol; 2) to share $a$ with Carol regardless of $s$. 3) To share $a$ when $s$ is 1, or 4) to share $a$  when $s$  is 0. 

\begin{thm}
    If Alice signals 1 when she is in a collaborative context with Carol and 0 when she is in a competitive context then Bob would share Alice's secret with Carol when Alice signals 1 and keep it when she signals 0. Hence, Bob has a strategy of Control towards Alice.
\end{thm}
\begin{proof}
From the analysis of the Game of Fiduciary Transfer we already know that keeping Alice's secret is harmful for Bob if Alice and Carol are in collaborative context and that sharing the secret is harmful to Bob when Alice and Carol are in competitive context. If Bob knows that Alice only signals 1 when she is in a collaborative context then his best sharing strategy is to share when the signal is 1 and to keep when it is 0. Since this strategy is also  the best secret sharing strategy for Alice, it is a payoff-optimal Nash equilibrium.
\end{proof}

\subsection{\label{subsec:Notification}Notification}

\begin{table}[htb]
\small
    \begin{subtable}[h]{0.24\textwidth}
    \centering
    \begin{tabular}{|c|c|c|c|c|}
        \cline{3-5}
        \multicolumn{1}{c}{} &  & \multicolumn{3}{c|}{{Carol}}\tabularnewline
        \cline{3-5}
        \multicolumn{1}{c}{} &  & {L} & {C} & {R}\tabularnewline
        \hline 
        \multirow{3}{*}{{\rotatebox{270}{Alice}}} 
            & {T} & {4,-8} & {0,0} & {-8,-8}\tabularnewline
        \cline{2-5}
            & {M} & {2,0} & {1,0} & {2,0}\tabularnewline
        \cline{2-5}
            & {B} & {-8,-8} & {0,0} & {4,-8}\tabularnewline
        \hline 
    \end{tabular}
    \caption{Alice v. Carol}
    \label{tab:Notify-Alice-v.-Carol}
    \end{subtable}
    \hfill
    \begin{subtable}[h]{0.24\textwidth}
    \centering
    \begin{tabular}{|c|c|c|c|c|}
        \cline{3-5}
        \multicolumn{1}{c}{} &  & \multicolumn{3}{c|}{{Bob}}\tabularnewline
        \cline{3-5}
        \multicolumn{1}{c}{} &  & {N} & {I} & {F}\tabularnewline
        \hline 
        \multirow{3}{*}{{\rotatebox{270}{Alice}}} 
            & {T} & {4,4} & {0,0} & {-8,-8}\tabularnewline
        \cline{2-5}
            & {M} & {1,0} & {1,0} & {1,0}\tabularnewline
        \cline{2-5}
            & {B} & {-8,-8} & {0,0} & {4,4}\tabularnewline
        \hline 
    \end{tabular}
    \caption{Alice v. Bob}
    \label{tab:Notify-Alice-v.-bob}
    \end{subtable}
    \hfill
    \begin{subtable}[h]{0.24\textwidth}
    \centering
    \begin{tabular}{|c|c|c|c|c|}
        \cline{3-5}
        \multicolumn{1}{c}{} &  & \multicolumn{3}{c|}{{Carol}}\tabularnewline
        \cline{3-5}
        \multicolumn{1}{c}{} &  & {L} & {C} & {R}\tabularnewline
        \hline 
        \multirow{3}{*}{{\rotatebox{270}{Bob}}} 
            & {N} & {0,0} & {0,0} & {0,0}\tabularnewline
        \cline{2-5}
            & {I} & {0,0} & {0,0} & {0,0}\tabularnewline
        \cline{2-5}
            & {F} & {0,0} & {0,0} & {0,0}\tabularnewline
        \hline 
    \end{tabular}
    \caption{Bob v. Carol - non collab.}
    \label{tab:Notify-Bob-v.-Carol-1}
    \end{subtable}
    \hfill
    \begin{subtable}[h]{0.24\textwidth}
    \centering
    \begin{tabular}{|c|c|c|c|c|}
        \cline{3-5}
        \multicolumn{1}{c}{} &  & \multicolumn{3}{c|}{{Carol}}\tabularnewline
        \cline{3-5}
        \multicolumn{1}{c}{} &  & {L} & {C} & {R}\tabularnewline
        \hline 
        \multirow{3}{*}{{\rotatebox{270}{Bob}}} 
            & {N} & -6,-6 & 0,0 & 6,6\tabularnewline
        \cline{2-5}
            & {I} & {0,0} & {0,0} & {0,0}\tabularnewline
        \cline{2-5}
            & {F} & 6,6 & 0,0 & -6,-6\tabularnewline
        \hline 
    \end{tabular}
    \caption{Bob v. Carol - collaborative}
    \label{tab:Notify-Bob-v.-Carol-2}
    \end{subtable}
    
\caption{Game of Notification}
\label{tab:Game-of-Notification}
\end{table}

A common real world example for notification is that in which a company is bought by another. 
When that happen, the context of the bought company and its new owner typically becomes collaborative. Such collaboration can be reinforced by sharing the bought company secrets with its owner. However, sometimes in order to collaborate with its owner the bought company needs to validate that its customers have a chance to alter their own strategy.

\begin{defn}
\label{def:notification}{[}Notification{]} Bob has a strategy of notification towards Alice if when she shares her secret a  with Bob, his strategy is to signal 1 to her if he chooses to share that secret with Carol and to signal 0 to her if he chooses to keep it.
\end{defn}

The Game of Notification in Table \ref{tab:Game-of-Notification} attempts to capture this dynamic by requiring that Alice choose her strategy a-priori, before any of the players knows if the context of Bob and Carol is collaborative. Alice strategy can still depend on a signal from Bob, but Alice would not be able to change her strategy in response to the choices of Bob and Carol. 

When Alice does not share her secret with Bob the only strategy which guarantees Alice a positive payoff is M and the only Nash equilibrium is $\left\langle M,I,C \right\rangle$ whose gain is $\left\langle 2,0,0\right\rangle$. By sharing her secret with Bob, Alice allows $\left\langle a?T:B,a?N:F,M \right\rangle$ whose gain is higher for her and for Bob, $\left\langle 5,4,0\right\rangle$. When Bob and Carol are in a non-collaborative context, that is the payoff-dominant Nash-equilibrium.

When Bob and Carol are in a collaborative context, or if they change their context to a collaborative one, they would prefer that Carol choose L when Bob chooses F and R when Bob chooses N. However, that is only a Nash equilibrium if Alice chooses M. Otherwise, e.g., in $\left\langle a?T:B,a?N:F,a?R:L\right\rangle$, the loss to Carol from choosing R when Alice chooses T is higher than Carol's gain from choosing R when Bob chooses N. 

Thus, there are two groups of Nash equilibria to the game when all players have Alice's secret: $\left\langle a?T:B,a?N:F, M \right\rangle$ with payoff $\left\langle 5,4,0 \right\rangle$  and $\left\langle M,a?N:F, a?R:L \right\rangle$ with payoff $\left\langle 3,6,6 \right\rangle$. Bob and Carol prefer the second and Alice prefer the first. However, Alice still prefers this second Nash equilibrium over $\left\langle M,I,C\right\rangle$ which is her best option if she does not share her secret with Bob.

Now consider the possibility that Bob signals to Alice $s=1$ if his context with Alice is collaborative and $s=0$  if non-collaborative. If Alice can trust that signal then her best strategy is $s?M:a?T:B$. Trust is critical because Bob's choice of what signal he sends comes after Alice has committed to the strategy and have shared her secret.
\begin{thm}
    If Alice shares her secret with Bob in the Notification Game then Bob has a strategy of notifying Alice if he shared her secret with Carol.
\end{thm}
\begin{proof}
    Bob would only share Alice's secret with Carol if his context with Carol is collaborative. If Bob shares the secret and then signals 0 then Alice strategy is $a?T:B$ and Carol would choose M. That would reduce Bob's payoff to 4 rather the 6 he could gain if he signalled 1. If Bob does not share Alice's secret and still signals 1 then Alice chooses M. That would reduce Bob's gain to 0 rather than the 4 he could have gained if he signalled 0.
\end{proof}

\section{\label{sec:mechanisms}Informational Norms as Mechanisms}
As stated earlier, an informational norm can be seen as a social mechanism
whose purpose is to make sure individuals adopt certain normative behaviors. 
In a game where players do not have information-normative strategies, a mechanism is an adaptations of the given game which cause players to choose an informational-normative strategy. This paper focuses on two types of mechanisms: Distributional mechanism in which the government can shift a player payoff to another player,or to itself, and communication mechanisms in which the government can obstruct the passing of some messages.

\subsection{Distributional mechanisms}
Two prevalent methods of payoff distribution in the commercial world are revenue-sharing and taxation\footnote{We consider any kind of \emph{monetary} punishment system as taxation for the purpose of this discussion. Since it allows players to make rational decisions which consider the probability of getting caught.}. In the former the data subject is paid for the use of her secret. In the latter, the government extracts some of the payoff from some players. 

\subsubsection{Information Ownership}

\begin{table}[htb]
\small
    \begin{subtable}[h]{0.24\textwidth}
    \centering
    \begin{tabular}{|c|c|c|c|c|}
        \cline{3-5}
        \multicolumn{1}{c}{} &  & \multicolumn{3}{c|}{{Carol}}\tabularnewline
        \cline{3-5}
        \multicolumn{1}{c}{} &  & {L} & {C} & {R}\tabularnewline
        \hline 
        \multirow{3}{*}{{\rotatebox{270}{Alice}}} 
            & {T} & {10,0} & {0,2} & {0,10}\tabularnewline
        \cline{2-5}
            & {M} & {2,0} & {2,2} & {2,0}\tabularnewline
        \cline{2-5}
            & {B} & {0,10} & {0,2} & {10,0}\tabularnewline
        \hline 
    \end{tabular}
    \caption{Alice v. Carol}
    \label{tab:Ownership-Alice-v.-Carol}
    \end{subtable}
    \hfill
    \begin{subtable}[h]{0.24\textwidth}
    \centering
    \begin{tabular}{|c|c|c|c|c|}
        \cline{3-5}
        \multicolumn{1}{c}{} &  & \multicolumn{3}{c|}{{Carol}}\tabularnewline
        \cline{3-5}
        \multicolumn{1}{c}{} &  & {L} & {C} & {R}\tabularnewline
        \hline 
        \multirow{3}{*}{{\rotatebox{270}{Bob}}} 
            & {N} & {0,0} & {0,0} & {0,0}\tabularnewline
        \cline{2-5}
            & {I} & {0,0} & {10,0} & {0,0}\tabularnewline
        \cline{2-5}
            & {F} & {0,0} & {0,0} & {0,0}\tabularnewline
        \hline 
    \end{tabular}
    \caption{Bob v. Carol}
    \label{tab:Ownership-Bob-v.-Carol}
    \end{subtable}
    \hfill
    \begin{subtable}[h]{0.24\textwidth}
    \centering    
    \begin{tabular}{|c|c|c|c|c|}
        \cline{3-5}
        \multicolumn{1}{c}{} &  & \multicolumn{3}{c|}{{Bob}}\tabularnewline
        \cline{3-5}
        \multicolumn{1}{c}{} &  & {N} & {I} & {F}\tabularnewline
        \hline 
        \multirow{3}{*}{{\rotatebox{270}{Alice}}} 
            & {T} & {2,2} & {0,0} & {0,0}\tabularnewline
        \cline{2-5}
            & {M} & {0,0} & {0,0} & {0,0}\tabularnewline
        \cline{2-5}
            & {B} & {0,0} & {0,0} & {2,2}\tabularnewline
        \hline 
    \end{tabular}
    \caption{Alice v. Bob - as given}
    \label{tab:Onership-Alice-v.-bob-Given}
    \end{subtable}
    \hfill
    \begin{subtable}[h]{0.24\textwidth}
    \centering
    \begin{tabular}{|c|c|c|c|c|}
        \cline{3-5}
        \multicolumn{1}{c}{} &  & \multicolumn{3}{c|}{{Bob}}\tabularnewline
        \cline{3-5}
        \multicolumn{1}{c}{} &  & {N} & {I} & {F}\tabularnewline
        \hline 
        \multirow{3}{*}{{\rotatebox{270}{Alice}}} 
            & {T} & {2,2} & {0,0} & {0,0}\tabularnewline
        \cline{2-5}
            & {M} & {0,0} & {5,-5} & {0,0}\tabularnewline
        \cline{2-5}
            & {B} & {0,0} & {0,0} & {2,2}\tabularnewline
        \hline 
    \end{tabular}
    \caption{Alice v. Bob - Modified}
    \label{tab:Onership-Alice-v.-bob-Modified}
    \end{subtable}
    
\caption{Game of Information Ownership}
\label{tab:Game-of-Ownership}
\end{table}

\begin{defn}
    {[}Information ownership{]} Bob has a strategy of recognising Alice's ownership of her secret if he chooses to transfer Alice some of his access payoff which resulted from sharing her secret with Carol.
 \end{defn}

In the Game of Information Ownership (Table \ref{tab:Game-of-Ownership}) Alice and Carol context is  a game of privacy. If Alice does not share her secret with Bob then the Nash equilibrium of the game is $\left\langle a?T:B,b?N:F,c?L:R\right\rangle$ with payoff $\left\langle 6,1,6 \right\rangle$. If Alice shares her secret with Bob then they can also choose $\left\langle a?T:B,a?N:F,c?L:R\right\rangle$ with payoff $\left\langle 7,2,6 \right\rangle$. However, Bob would much rather Alice would choose M because the strategy profile $\left\langle M,I,C\right\rangle$ gains him 10 from Carol. 

If Alice shares her secret with Bob then he can force $\left\langle M,I,C\right\rangle$  by sharing Alice's secret with Carol. 
As the game is given, Alice see her gain reduced from 6 if she does not share her secret with Bob, to 2 if she does share her secret with Bob (and he shares it with Carol). Therefore, as the game is given, Alice would not share her secret with Bob.

If Alice and Bob could adopt a mechanism in which Bob transfers a payment of 5 to Alice if in their context they choose M and I (Table \ref{tab:Onership-Alice-v.-bob-Modified}) then Alice's calculation would change. Now, in $\left\langle M,I,C\right\rangle$  her gain is 7 whereas in  $\left\langle a?T:B,b?N:F,c?L:R\right\rangle$  it is 6. Alice no longer minds that Bob might share her secret with Carol. Since Bob increases his gain in $\left\langle M,I,C\right\rangle$ to 5 comparing to 1 in  $\left\langle a?T:B,b?N:F,c?L:R\right\rangle$, the mechanism works better for Bob as well. Last, since the total gain in  $\left\langle M,I,C\right\rangle$ is 12, whereas in  $\left\langle a?T:B,b?N:F,c?L:R\right\rangle$ it is 11, the mechanism is ethical from a utilitarian point of view.

\subsubsection{Taxation}
The same game which exemplifies Information Ownership can also present the value of taxation. Consider if in the game in Table \ref{tab:Game-of-Ownership}, the government placed a flat tax of 5 on Alice if she chooses anything but M. That would cause Alice to prefer $\left\langle M,I,C \right\rangle$ over $\left\langle a?T:B,b?N:F,c?L:R \right\rangle$  because in the former Alice's gain is 2 which in the later it is 1. Since the total gain of $\left\langle M,I,C \right\rangle$ is higher than that of  $\left\langle a?T:B,b?N:F,c?L:R \right\rangle$, that tax is also ethical from a utilitarian point of view\footnote{Similar examples can be brought in which the tax also increase the total after-tax payoff of the players.}.  

\subsection{Communication mechanisms}
Secret sharing was presented so far as optimal communication: Whenever a source player decides to share a secret the destination player immediately knows the value of that secret. There are many ways in which such perfect communication can be degraded. Firstly, the channel can be made interactive, requiring that the destination be open to receiving the secret or else it does not matter if the source player chooses to share it. Secondly, the channel can be made noisy, such that when the origin shares a secret $s$  the destination receives $\tilde{s}$  which is only equal to $s$ in some probability. Thirdly, the bandwidth of the channel can be limited such that no more than $k$  secrets can be shared even if the origin has $n$ secrets she wishes to share. 

\subsubsection{Interactive channel mechanism} A real life example of making a  channel interactive is the "fruit of the poisonous tree" doctrine  \cite{1920silverthorne}. According to that doctrine, the court (recipient)  can choose not to legally know some facts  which the prosecution (sender) decided to share about the defendant (subject). The doctrine is justified when ignoring those facts serves the greater social good of discouraging law enforcement from unlawful behavior.

\begin{table}[htb]
\small
    \begin{subtable}[h]{0.36\textwidth}
    \centering
    \begin{tabular}{|c|c|c|c|c|}
        \cline{3-5} 
        \multicolumn{1}{c}{} &  & \multicolumn{3}{c|}{{Carol}}\tabularnewline
        \cline{3-5} 
        \multicolumn{1}{c}{} &  & {L} & {C} & {R}\tabularnewline
        \hline 
        \multirow{3}{*}{{\rotatebox{270}{Alice}}} & {T} & {0,16} & {0,4} & {16,0}\tabularnewline
        \cline{2-5} 
            & {M} & {4,0} & {4,4} & {4,0}\tabularnewline
        \cline{2-5} 
        & {B} & {16,0} & {0,4} & {0,16}\tabularnewline
        \hline 
    \end{tabular}
    \caption{Alice v. Carol}
    \label{tab:poisonous-Alice-v.-Carol}
    \end{subtable}
    \hfill
    \begin{subtable}[h]{0.27\textwidth}
    \centering
    \begin{tabular}{|c|c|c|c|c|}
        \cline{3-5}
        \multicolumn{1}{c}{} &  & \multicolumn{3}{c|}{{Bob}}\tabularnewline
        \cline{3-5}
        \multicolumn{1}{c}{} &  & {N} & {I} & {F}\tabularnewline
        \hline 
        \multirow{3}{*}{{\rotatebox{270}{Alice}}} 
            & {T} & {2,2} & {0,0} & {0,0}\tabularnewline
        \cline{2-5}
            & {M} & {0,0} & {2,2} & {0,0}\tabularnewline
        \cline{2-5}
            & {B} & {0,0} & {0,0} & {2,2}\tabularnewline
        \hline 
    \end{tabular}
    \caption{Alice v. Bob}
    \label{tab:poisonous-Alice-v.-Bob}
    \end{subtable}
    \hfill
    \begin{subtable}[h]{0.27\textwidth}
    \centering
    \begin{tabular}{|c|c|c|c|c|}
        \cline{3-5}
        \multicolumn{1}{c}{} &  & \multicolumn{3}{c|}{{Carol}}\tabularnewline
        \cline{3-5}
        \multicolumn{1}{c}{} &  & {L} & {C} & {R}\tabularnewline
        \hline 
        \multirow{3}{*}{{\rotatebox{270}{Bob}}} 
        & {N} & {0,8} & {0,2} & {8,0}\tabularnewline
        \cline{2-5}
            & {I} & {2,0} & {5,2} & {2,0}\tabularnewline
        \cline{2-5}
            & {F} & {8,0} & {0,2} & {0,8}\tabularnewline
        \hline 
    \end{tabular}
    \caption{Bob v. Carol}
    \label{tab:poisonous-Bob-v.-Carol}
    \end{subtable}
\caption{Interactive channel mechanism}
\label{tab:Game-of-poisonous}
\end{table}

In  the interactive channel mechanism example (Table \ref{tab:Game-of-poisonous}) Bob prefers the Nash equilibrium $\left\langle M, I, C\right\rangle$ which pays $\left\langle 6, 7, 6\right\rangle$  over the Nash equilibrium $\left\langle a?T:B, b?N:F,c?L:R \right\rangle$  which pays $\left\langle 9, 5, 12\right\rangle$  and over  $\left\langle a?T:B, a?N:F,c?L:R \right\rangle$  which pays $\left\langle 10, 6, 12\right\rangle$ . If Bob learns Alice's secret then he can force the Nash equilibrium he prefers by sharing it with Carol. Knowing that, Alice would never choose to share her secret with Bob.  But Bob might still (illicitly) observe it.

Consider what would happen if the government enforces a communication mechanism by which Carol can choose if she learns secrets Bob wants to share with her. Since Carol stands only to loose if she knows Alice's secret, she will choose to not learn it from Bob. When that is Carol's strategy Alice can  safely share her secret with Bob, making  $\left\langle a?T:B, a?N:F,c?L:R \right\rangle$  the payoff-optimal Nash equilibrium. Since the introduction of the mechanism increased the total payoff from 26 to 28, it is ethical from a utilitarian point of view.

\subsubsection{Noisy channel}

\begin{table}[htb]
\small
    \begin{subtable}[h]{0.36\textwidth}
    \centering
    \begin{tabular}{|c|c|c|c|c|}
        \cline{3-5} 
        \multicolumn{1}{c}{} &  & \multicolumn{3}{c|}{{Carol}}\tabularnewline
        \cline{3-5} 
        \multicolumn{1}{c}{} &  & {L} & {C} & {R}\tabularnewline
        \hline 
        \multirow{3}{*}{{\rotatebox{270}{Alice}}} & {T} & {0,16} & {0,4} & {16,0}\tabularnewline
        \cline{2-5} 
            & {M} & {4,0} & {4,4} & {4,0}\tabularnewline
        \cline{2-5} 
        & {B} & {16,0} & {0,4} & {0,16}\tabularnewline
        \hline 
    \end{tabular}
    \caption{Alice v. Carol}
    \label{tab:noisy-Alice-v.-Carol}
    \end{subtable}
    \hfill
    \begin{subtable}[h]{0.27\textwidth}
    \centering
    \begin{tabular}{|c|c|c|c|c|}
        \cline{3-5}
        \multicolumn{1}{c}{} &  & \multicolumn{3}{c|}{{Bob}}\tabularnewline
        \cline{3-5}
        \multicolumn{1}{c}{} &  & {N} & {I} & {F}\tabularnewline
        \hline 
        \multirow{3}{*}{{\rotatebox{270}{Alice}}} 
            & {T} & {2,2} & {0,0} & {0,0}\tabularnewline
        \cline{2-5}
            & {M} & {0,0} & {0,0} & {0,0}\tabularnewline
        \cline{2-5}
            & {B} & {0,0} & {0,0} & {2,2}\tabularnewline
        \hline 
    \end{tabular}
    \caption{Alice v. Bob}
    \label{tab:noisy-Alice-v.-Bob}
    \end{subtable}
    \hfill
    \begin{subtable}[h]{0.27\textwidth}
    \centering
    \begin{tabular}{|c|c|c|c|c|}
        \cline{3-5}
        \multicolumn{1}{c}{} &  & \multicolumn{3}{c|}{{Carol}}\tabularnewline
        \cline{3-5}
        \multicolumn{1}{c}{} &  & {L} & {C} & {R}\tabularnewline
        \hline 
        \multirow{3}{*}{{\rotatebox{270}{Bob}}} 
        & {N} & {6,6} & {0,0} & {-6,-6}\tabularnewline
        \cline{2-5}
            & {I} & {0,0} & {0,0} & {0,0}\tabularnewline
        \cline{2-5}
            & {F} & {-6,-6} & {0,0} & {6,6}\tabularnewline
        \hline 
    \end{tabular}
    \caption{Bob v. Carol}
    \label{tab:noisy-Bob-v.-Carol}
    \end{subtable}
\caption{Noisy channel mechanism}
\label{tab:Game-of-noisy}
\end{table}

Differential privacy \cite{dwork2008differential} is, possibly the most well known example of a noisy channel mechanism. In the Noisy channel mechanism example (Table \ref{tab:Game-of-noisy}) Alice can increase her expected payoff from 9 in the payoff-optimal Nash equilibrium $\left\langle a?T:B,b?N:F,c?L:R\right\rangle$  to 10 in the payoff-optimal Nash equilibrium  $\left\langle a?T:B,a?N:F,c?L:R\right\rangle$ by sharing her secret with Bob. Bob, in turn, has no motivation to share Alice's secret with Carol because that would force the Nash equilibrium  $\left\langle M,I,C\right\rangle$  and reduce his payoff from 2 to 0.

Assume the government enforces a noisy channel mechanism in which Bob can share not $a$ but $\tilde{a}$ such that the chances that $\tilde{a} = a$ are $\frac{1}{2}+\delta$. Carol can still choose the strategy $\tilde{a}?L:R$, rather than $a?L:R$ in the given game. The payoff of the strategy profile $\left\langle a?T:B,a?N:F,\tilde{a}?L:R\right\rangle$ is  $\left\langle 10-16\delta,2+12\delta,8+28\delta\right\rangle$. That payoff is higher for Bob and Carol than the payoff of $\left\langle a?T:B, a?N:F,c?L:R\right\rangle$, and  so long as $10-16\delta > 4$ it is also higher for Alice than her alternative gain if she chooses M. Hence, it is a payoff-optimal Nash equilibrium. The total payoff of this Nash equilibrium, $20+24\delta$, is higher of the total payoff of 18 that  $\left\langle a?T:B,a?N:F,c?L:R\right\rangle$  delivers. Therefore, the noisy channel mechanism is ethical from a utilitarian point of view.

\subsubsection{Bandwidth limitation} Last, consider the same game which is described in Table \ref{tab:Game-of-noisy} but assume now Bob and Carol repeatedly play it with $k$ different Alice-s. Each Alice has her own secret $a_i$ and every player can choose a different  and can independently choose her own strategy. Both Bob and Carol must choose one strategy each which simultaneously applies to all of the contexts.

As in the noisy channel mechanism, each Alice can increase her expected payoff from 9 in the payoff-optimal Nash equilibria $\left\langle a_i?T:B,b?N:F,c?L:R\right\rangle$  to 10 in the payoff-optimal Nash equilibria  $\left\langle a_i?T:B,a_i?N:F,c?L:R\right\rangle$ by sharing her secret with Bob. Bob, however, can share those secrets over with Carol and force the Nash equilibria  $\left\langle M,a_i?N:F,a_i?L:R\right\rangle$ which pays Bob $6k$ rather than $2k$. Since every Alice's payoff in $\left\langle M,a_i?N:F,a_i?L:R\right\rangle$  is just 4, she will not share $a_i$ with Bob in the first place.

Assume that the government implements a communication mechanism which only allows Bob to share one bit with Carol. Bob can select that  bit to be any of the secrets, or any function of  the secrets. Specifically consider Bob's strategy of sharing $f$ such that $f=0$ if less than a fraction $\alpha$ of the $a_i$ are 1, $f=1$ if more than $1-\alpha$ are 1, and $f=b$, Bob's secret, if the fraction is between $\alpha$ and $1-\alpha$.

With that strategy, the strategy profile $\left\langle a_i?T:B,a_i?N:F,f?L:R\right\rangle$  is possible. Central limit theorem dictates that for a large enough k the chances that a fraction of less than $\alpha$ or more than $1-\alpha$ of the $a_i$ is 1 is equal to the probability that a normally distributed variable $N\left(0,\frac{1}{4}\right)$ exceeds the range  $\left(\alpha,1-\alpha\right)$ . If $\alpha$ is taken to be $\frac{1}{4}$ then with probability of approximately 68\% (one $\sigma$ )  $f=b$. Thus, the expected payoff of  every Alice in $\left\langle a_i?T:B,a_i?N:F,f?L:R\right\rangle$  is higher than 68\% of 10, which is higher than her alternative payoff if she chooses M. Since that strategy profile also increases Bob and Carol's payoff, it follows that if $k$ is sufficiently large and $\alpha$ sufficiently small then $\left\langle a_i?T:B,a_i?N:F,f?L:R\right\rangle$ is a Nash equilibrium. 
As in the noisy channel mechanism, the total expected payoff of this bandwidth limiting mechanism is higher than that of the original game, which means it is ethical from a utilitarian point of view.

\section{\label{sec:related} Related Work}

Work related to this paper  can largely be divided between that which
aims to explain what privacy means and that which aims to explain how
privacy can be retained. The first kind is more widely practice by
legal scholars \cite{brandeis1890right, gavison1980privacy,soloveUnderstandingPrivacy}, sociologists \cite{boyd2012networked,marwick2018understanding}, economists \cite{GRADWOHL2017293} and a few exceptional scholars who thrive in more than one of those fields \cite{posner1977right,nissenbaumCI,ohm2009broken}.  The second kind was mostly developed by technologists in the Computer and Data Sciences.

\subsection{Rigorous definitions of privacy}

A key objective of the philosophy of privacy is to provide a rigorous and useful
definition of  privacy.
Gavison \cite{gavison1980privacy} described privacy as the meeting point of three vectors (knowledge,
ability to affect, and attention to a person). Posner \cite{posner1977right} placed privacy
in a utilitarian economic framework, famously attesting that government
regulation of privacy is an unnecessary intervention in a free market
for information. While this claim was objected for almost immediately \cite{baker1977posner} it was Kadane et al. \cite{kadane2008ignorance} who finally refuted that claim  by quantitatively showing that 
privacy does make perfect utilitarian sense in a game theoretic framework, if not in a Bayesian efficient market.

One of the most important contribution to rigorously defining privacy was the development
of  Contextual Integrity by Nissenbaum \cite{nissenbaumCI,CIBook}. Nissenbaum suggested to inspect
privacy as the interaction of three social actors -- subject, sender
and recipient -- in two separate social contexts: A social exchange
in the first context leaves the sender with information about the subject.
Then, privacy norms regulate the transfer of that information from
the sender to the recipient in another social context.
This work extends contextual integrity by placing it in a
strict game theoretical setting. In such settings  actors are rational
decision makers who act to maximize their gains in an environment
which contains secrets.

Previous work on game theoretic privacy has built on Kadane et al. and has shown that privacy, like secrecy \cite{dighe2009secrecy}, can be described as the strategy of players with respect to secrets.
Anonymous \cite{wolff2015emergent} has shown that when one player has access to another player's secret, the first might still choose to respect the other's privacy. Thus, a privacy norm can emerge as the strategy in certain games. 
A similar result was later presented by Ulusoy and Pinar \cite{ulusoy2019emergent}, in a multi-player scenario. Still, Ulusoy and Pinar do not reference their result to contextual integrity, which is the main contribution of this work.

The impact of privacy on games was studied by Gradwohl and Reingold \cite{GRADWOHL2010602} and Gradwohl and Smorodinsky \cite{GRADWOHL2017293}. The first work has shown that in multiplayer simultaneous games players would rather not expose their secret type. In the notation proposed here, this is related more to the concept of secrecy (Def. \ref{def:secrecy}) than it is to privacy. The second paper begin by assuming players desire to not be identified. The authors then draw conclusion on the societal impact (pooling) of such games. This, however, assumes a privacy preference rather than explain why such preference may exist in the first place, as this paper does.

\subsection{Privacy preserving mechanisms}
The leading paradigm of privacy preserving mechanism, initially proposed by Warner \cite{warnerRandomizedResponse}, is that of a data collector who wishes to publish a statistics of the data of multiple subjects. Those subjects are concerned of the implications of the publication. The role of the mechanism is to contain that risk.

Privacy preserving mechanisms can be divided to those which are operated by the data subject, without need to trust the collector, and those operated by a trusted collector. Warner's original work falls into the first category, as do most of the Secure Multiparty Computation \cite{lindell2000privacy,10.1007/978-3-540-28628-8_32}. Kantarcioundefinedlu et al. \cite{10.1145/1014052.1014126} and Anonymous \cite{wolff2004kttp} have independently identified that the outcome of the computation, rather than its security, might concern the data subjects. Anonymous offered composition of security and $k$-anonymity as a solution \cite{wolff2004kttp}. 

Research on trusted collector mechanisms can be divided to noisy channel mechanisms (a.k.a., data perturbation) and mechanisms which rely on bandwidth limitation (aggregation). The first kind was initially suggested by Agrawal and Srikant \cite{agrawal2000privacy} without proper analysis of the impact of noise on the players. Evfimievski et al. \cite{evfimievski2003limiting} were the first to try and quantify the impact of noise on the certainty of the recipient. Dwork, on her own \cite{dwork2006differential} and with co-authors \cite{dwork2006edDiffPriv}, presented a full analysis of the impact of noise on the information transfer from the collector to the recipient. This included a proof of the infeasibility of zero leakage, a definition of the worst case model of Differential Privacy, and first algorithms. Hundreds of studies, which cannot reasonably be surveyed here, have since validated the usefulness of differential privacy.

Differential Privacy stops shy of inspecting the impact of data leakage on the data subject. That analysis is partially fulfilled by Gilboa-Freedman and Smorodinsky \cite{gilboa2020behavioral} who have shown that this impact can be different in different types of games. This paper follows a similar path. Gilboa-Freedman and Smorodinsky  go beyond this work in inspecting the equivalence and non-equivalence of different privacy mechanism. However, they focus primarily on preserving near-confidentiality whereas this work expands the analysis to other informative-norms. 

Last, data aggregation was first proposed by Sweeney and Samarati \cite{samarati1998protecting} as a way of protecting data subjects from specific attacks by the recipient. Machanavajjhala at el. \cite{1617392} developed a more elaborate concept of aggregation. Attempts to quantify the impact of $k$-anonymity have mostly drawn parallels to $\epsilon$-Differential Privacy \cite{li2011provably,danezis2013measuring, domingo2015t}. This paper proposes looking at multiparty games, rather than on noisy channels, as the adequate model in which the impact of $k$-anonymity may best be quantified. 

\section{\label{sec:discussion}Discussion}

This paper presented a game theoretic transcription of Nissenbaum's contextual integrity model. 
Game theory allows explaining the existence of privacy norms in terms of players' payoff and social welfare. This utilitarian analysis of privacy is especially adequate when norms are set by a revenue maximizing corporate and their willing customers. 

One other benefit of a game theoretic model is that it allows analysis of privacy related situations in the sense of sufficiency and equivalence. It allows answering questions such as: How much information can a company share with advertisers before customers start changing their data sharing behavior? When is the threat of loosing one's job as effective as a technology which limits ones access to individuals' data? How does one compare the risk of being identified on-line to other risks such as the risk of loosing a key elections? 

In terms of future research, we observe that privacy preserving technology which were developed in the last 30 years have focused on data subjects' control of their information and on data aggregator confidentiality. We hope that  mathematical definitions of fiduciary transfer and of information ownership can lead to the development of technologies implementing those transmission principles as well.

\newpage
\bibliographystyle{ACM-Reference-Format}
\bibliography{ref}


\begin{thebibliography}{39}


\ifx \showCODEN    \undefined \def \showCODEN     #1{\unskip}     \fi
\ifx \showDOI      \undefined \def \showDOI       #1{#1}\fi
\ifx \showISBNx    \undefined \def \showISBNx     #1{\unskip}     \fi
\ifx \showISBNxiii \undefined \def \showISBNxiii  #1{\unskip}     \fi
\ifx \showISSN     \undefined \def \showISSN      #1{\unskip}     \fi
\ifx \showLCCN     \undefined \def \showLCCN      #1{\unskip}     \fi
\ifx \shownote     \undefined \def \shownote      #1{#1}          \fi
\ifx \showarticletitle \undefined \def \showarticletitle #1{#1}   \fi
\ifx \showURL      \undefined \def \showURL       {\relax}        \fi
\providecommand\bibfield[2]{#2}
\providecommand\bibinfo[2]{#2}
\providecommand\natexlab[1]{#1}
\providecommand\showeprint[2][]{arXiv:#2}

\bibitem[192(1920)]%
        {1920silverthorne}
 \bibinfo{year}{1920}\natexlab{}.
\newblock \showarticletitle{Silverthorne Lumber Co. v. United States}.
\newblock  \bibinfo{volume}{251}, \bibinfo{number}{No. 358}
  (\bibinfo{year}{1920}).
\newblock


\bibitem[isr(1992)]%
        {israel_law_review_1992}
 \bibinfo{year}{1992}\natexlab{}.
\newblock \bibinfo{title}{Basic Law: Human Dignity and Freedom}.
\newblock , \bibinfo{numpages}{248–249}~pages.
\newblock
\urldef\tempurl%
\url{https://doi.org/10.1017/S0021223700010943}
\showDOI{\tempurl}


\bibitem[Agrawal and Srikant(2000)]%
        {agrawal2000privacy}
\bibfield{author}{\bibinfo{person}{Rakesh Agrawal} {and}
  \bibinfo{person}{Ramakrishnan Srikant}.} \bibinfo{year}{2000}\natexlab{}.
\newblock \showarticletitle{Privacy-preserving data mining}. In
  \bibinfo{booktitle}{\emph{Proceedings of the 2000 ACM SIGMOD international
  conference on Management of data}}. \bibinfo{pages}{439--450}.
\newblock


\bibitem[Anonymous( a)]%
        {wolff2015emergent}
\bibfield{author}{\bibinfo{person}{Anonymous}.} \bibinfo{year}{-}\natexlab{a}.
\newblock \showarticletitle{-}.
\newblock \bibinfo{journal}{\emph{-}} \bibinfo{volume}{-}, \bibinfo{number}{-}
  (\bibinfo{year}{-}), \bibinfo{pages}{X--X}.
\newblock


\bibitem[Anonymous( b)]%
        {wolff2004kttp}
\bibfield{author}{\bibinfo{person}{Anonymous}.} \bibinfo{year}{-}\natexlab{b}.
\newblock \showarticletitle{-}.
\newblock \bibinfo{journal}{\emph{-}}  \bibinfo{volume}{-},
  \bibinfo{pages}{X--X}.
\newblock


\bibitem[Baker(1977)]%
        {baker1977posner}
\bibfield{author}{\bibinfo{person}{C~Edwin Baker}.}
  \bibinfo{year}{1977}\natexlab{}.
\newblock \showarticletitle{Posner's Privacy Mystery and the Failure of
  Economic Analysis of Law}.
\newblock \bibinfo{journal}{\emph{Ga. L. Rev.}}  \bibinfo{volume}{12}
  (\bibinfo{year}{1977}), \bibinfo{pages}{475}.
\newblock


\bibitem[Boyd(2012)]%
        {boyd2012networked}
\bibfield{author}{\bibinfo{person}{Danah Boyd}.}
  \bibinfo{year}{2012}\natexlab{}.
\newblock \showarticletitle{Networked privacy}.
\newblock \bibinfo{journal}{\emph{Surveillance \& society}}
  \bibinfo{volume}{10}, \bibinfo{number}{3/4} (\bibinfo{year}{2012}),
  \bibinfo{pages}{348}.
\newblock


\bibitem[Brandeis and Warren(1890)]%
        {brandeis1890right}
\bibfield{author}{\bibinfo{person}{Louis Brandeis} {and}
  \bibinfo{person}{Samuel Warren}.} \bibinfo{year}{1890}\natexlab{}.
\newblock \showarticletitle{The right to privacy}.
\newblock \bibinfo{journal}{\emph{Harvard law review}} \bibinfo{volume}{4},
  \bibinfo{number}{5} (\bibinfo{year}{1890}), \bibinfo{pages}{193--220}.
\newblock


\bibitem[Danezis(2013)]%
        {danezis2013measuring}
\bibfield{author}{\bibinfo{person}{George Danezis}.}
  \bibinfo{year}{2013}\natexlab{}.
\newblock \showarticletitle{Measuring anonymity: a few thoughts and a
  differentially private bound}. In \bibinfo{booktitle}{\emph{Proceedings of
  the DIMACS Workshop on Measuring Anonymity}}. \bibinfo{pages}{26}.
\newblock


\bibitem[Dighe et~al\mbox{.}(2009)]%
        {dighe2009secrecy}
\bibfield{author}{\bibinfo{person}{Nikhil~S Dighe}, \bibinfo{person}{Jun
  Zhuang}, {and} \bibinfo{person}{Vicki~M Bier}.}
  \bibinfo{year}{2009}\natexlab{}.
\newblock \showarticletitle{Secrecy in Defensive Allocations as a Strategy for
  achieving more Cost-effective Attacker Deterrence}.
\newblock \bibinfo{journal}{\emph{International Journal of Performability
  Engineering}} \bibinfo{volume}{5}, \bibinfo{number}{1}
  (\bibinfo{year}{2009}), \bibinfo{pages}{31}.
\newblock


\bibitem[Dollinger(2014)]%
        {dollinger2014judicial}
\bibfield{author}{\bibinfo{person}{Richard~A Dollinger}.}
  \bibinfo{year}{2014}\natexlab{}.
\newblock \showarticletitle{Judicial Ethics: The Obligation to Report Tax
  Evasion in Support Cases}.
\newblock \bibinfo{journal}{\emph{Journal of the American Academy of
  Matrimonial Lawyers}}  \bibinfo{volume}{27} (\bibinfo{year}{2014}),
  \bibinfo{pages}{1}.
\newblock


\bibitem[Domingo-Ferrer and Soria-Comas(2015)]%
        {domingo2015t}
\bibfield{author}{\bibinfo{person}{Josep Domingo-Ferrer} {and}
  \bibinfo{person}{Jordi Soria-Comas}.} \bibinfo{year}{2015}\natexlab{}.
\newblock \showarticletitle{From t-closeness to differential privacy and vice
  versa in data anonymization}.
\newblock \bibinfo{journal}{\emph{Knowledge-Based Systems}}
  \bibinfo{volume}{74} (\bibinfo{year}{2015}), \bibinfo{pages}{151--158}.
\newblock


\bibitem[Dwork(2006)]%
        {dwork2006differential}
\bibfield{author}{\bibinfo{person}{Cynthia Dwork}.}
  \bibinfo{year}{2006}\natexlab{}.
\newblock \showarticletitle{Differential privacy}. In
  \bibinfo{booktitle}{\emph{International colloquium on automata, languages,
  and programming}}. Springer, \bibinfo{pages}{1--12}.
\newblock


\bibitem[Dwork(2008)]%
        {dwork2008differential}
\bibfield{author}{\bibinfo{person}{Cynthia Dwork}.}
  \bibinfo{year}{2008}\natexlab{}.
\newblock \showarticletitle{Differential privacy: A survey of results}. In
  \bibinfo{booktitle}{\emph{Theory and Applications of Models of Computation:
  5th International Conference, TAMC 2008, Xian, China, April 25-29, 2008.
  Proceedings 5}}. Springer, \bibinfo{pages}{1--19}.
\newblock


\bibitem[Dwork et~al\mbox{.}(2006)]%
        {dwork2006edDiffPriv}
\bibfield{author}{\bibinfo{person}{Cynthia Dwork}, \bibinfo{person}{Krishnaram
  Kenthapadi}, \bibinfo{person}{Frank McSherry}, \bibinfo{person}{Ilya
  Mironov}, {and} \bibinfo{person}{Moni Naor}.}
  \bibinfo{year}{2006}\natexlab{}.
\newblock \showarticletitle{Our data, ourselves: Privacy via distributed noise
  generation}. In \bibinfo{booktitle}{\emph{Advances in Cryptology-EUROCRYPT
  2006: 24th Annual International Conference on the Theory and Applications of
  Cryptographic Techniques, St. Petersburg, Russia, May 28-June 1, 2006.
  Proceedings 25}}. Springer, \bibinfo{pages}{486--503}.
\newblock


\bibitem[Dwork and Nissim(2004)]%
        {10.1007/978-3-540-28628-8_32}
\bibfield{author}{\bibinfo{person}{Cynthia Dwork} {and} \bibinfo{person}{Kobbi
  Nissim}.} \bibinfo{year}{2004}\natexlab{}.
\newblock \showarticletitle{Privacy-Preserving Datamining on Vertically
  Partitioned Databases}. In \bibinfo{booktitle}{\emph{Advances in Cryptology
  -- CRYPTO 2004}}, \bibfield{editor}{\bibinfo{person}{Matt Franklin}} (Ed.).
  \bibinfo{publisher}{Springer Berlin Heidelberg}, \bibinfo{address}{Berlin,
  Heidelberg}, \bibinfo{pages}{528--544}.
\newblock
\showISBNx{978-3-540-28628-8}


\bibitem[Evfimievski et~al\mbox{.}(2003)]%
        {evfimievski2003limiting}
\bibfield{author}{\bibinfo{person}{Alexandre Evfimievski},
  \bibinfo{person}{Johannes Gehrke}, {and} \bibinfo{person}{Ramakrishnan
  Srikant}.} \bibinfo{year}{2003}\natexlab{}.
\newblock \showarticletitle{Limiting privacy breaches in privacy preserving
  data mining}. In \bibinfo{booktitle}{\emph{Proceedings of the twenty-second
  ACM SIGMOD-SIGACT-SIGART symposium on Principles of database systems}}.
  \bibinfo{pages}{211--222}.
\newblock


\bibitem[Gavison(1980)]%
        {gavison1980privacy}
\bibfield{author}{\bibinfo{person}{Ruth Gavison}.}
  \bibinfo{year}{1980}\natexlab{}.
\newblock \showarticletitle{Privacy and the Limits of Law}.
\newblock \bibinfo{journal}{\emph{The Yale law journal}} \bibinfo{volume}{89},
  \bibinfo{number}{3} (\bibinfo{year}{1980}), \bibinfo{pages}{421--471}.
\newblock


\bibitem[Gilboa-Freedman and Smorodinsky(2020)]%
        {gilboa2020behavioral}
\bibfield{author}{\bibinfo{person}{Gail Gilboa-Freedman} {and}
  \bibinfo{person}{Rann Smorodinsky}.} \bibinfo{year}{2020}\natexlab{}.
\newblock \showarticletitle{On the behavioral implications of differential
  privacy}.
\newblock \bibinfo{journal}{\emph{Theoretical Computer Science}}
  \bibinfo{volume}{841} (\bibinfo{year}{2020}), \bibinfo{pages}{84--93}.
\newblock


\bibitem[Gradwohl and Reingold(2010)]%
        {GRADWOHL2010602}
\bibfield{author}{\bibinfo{person}{Ronen Gradwohl} {and} \bibinfo{person}{Omer
  Reingold}.} \bibinfo{year}{2010}\natexlab{}.
\newblock \showarticletitle{Partial exposure in large games}.
\newblock \bibinfo{journal}{\emph{Games and Economic Behavior}}
  \bibinfo{volume}{68}, \bibinfo{number}{2} (\bibinfo{year}{2010}),
  \bibinfo{pages}{602--613}.
\newblock
\showISSN{0899-8256}
\urldef\tempurl%
\url{https://doi.org/10.1016/j.geb.2009.09.006}
\showDOI{\tempurl}


\bibitem[Gradwohl and Smorodinsky(2017)]%
        {GRADWOHL2017293}
\bibfield{author}{\bibinfo{person}{Ronen Gradwohl} {and} \bibinfo{person}{Rann
  Smorodinsky}.} \bibinfo{year}{2017}\natexlab{}.
\newblock \showarticletitle{Perception games and privacy}.
\newblock \bibinfo{journal}{\emph{Games and Economic Behavior}}
  \bibinfo{volume}{104} (\bibinfo{year}{2017}), \bibinfo{pages}{293--308}.
\newblock
\showISSN{0899-8256}
\urldef\tempurl%
\url{https://doi.org/10.1016/j.geb.2017.04.006}
\showDOI{\tempurl}


\bibitem[Helen(2010)]%
        {CIBook}
\bibfield{author}{\bibinfo{person}{Nissenbaum Helen}.}
  \bibinfo{year}{2010}\natexlab{}.
\newblock \bibinfo{booktitle}{\emph{Privacy in Context : Technology, Policy,
  and the Integrity of Social Life.}}
\newblock \bibinfo{publisher}{Stanford Law Books}.
\newblock
\showISBNx{9780804752374}


\bibitem[Henkin(1974)]%
        {henkin1974privacy}
\bibfield{author}{\bibinfo{person}{Louis Henkin}.}
  \bibinfo{year}{1974}\natexlab{}.
\newblock \showarticletitle{Privacy and autonomy}.
\newblock \bibinfo{journal}{\emph{Columbia Law Review}}  \bibinfo{volume}{74}
  (\bibinfo{year}{1974}), \bibinfo{pages}{1410}.
\newblock


\bibitem[Kadane et~al\mbox{.}(2008)]%
        {kadane2008ignorance}
\bibfield{author}{\bibinfo{person}{Joseph~B Kadane}, \bibinfo{person}{Mark
  Schervish}, {and} \bibinfo{person}{Teddy Seidenfeld}.}
  \bibinfo{year}{2008}\natexlab{}.
\newblock \showarticletitle{Is ignorance bliss?}
\newblock \bibinfo{journal}{\emph{The Journal of Philosophy}}
  \bibinfo{volume}{105}, \bibinfo{number}{1} (\bibinfo{year}{2008}),
  \bibinfo{pages}{5--36}.
\newblock


\bibitem[Kantarcioundefinedlu et~al\mbox{.}(2004)]%
        {10.1145/1014052.1014126}
\bibfield{author}{\bibinfo{person}{Murat Kantarcioundefinedlu},
  \bibinfo{person}{Jiashun Jin}, {and} \bibinfo{person}{Chris Clifton}.}
  \bibinfo{year}{2004}\natexlab{}.
\newblock \showarticletitle{When do data mining results violate privacy?}. In
  \bibinfo{booktitle}{\emph{Proceedings of the Tenth ACM SIGKDD International
  Conference on Knowledge Discovery and Data Mining}} (Seattle, WA, USA)
  \emph{(\bibinfo{series}{KDD '04})}. \bibinfo{publisher}{Association for
  Computing Machinery}, \bibinfo{address}{New York, NY, USA},
  \bibinfo{pages}{599–604}.
\newblock
\showISBNx{1581138881}
\urldef\tempurl%
\url{https://doi.org/10.1145/1014052.1014126}
\showDOI{\tempurl}


\bibitem[Li et~al\mbox{.}(2011)]%
        {li2011provably}
\bibfield{author}{\bibinfo{person}{Ninghui Li}, \bibinfo{person}{Wahbeh~H
  Qardaji}, {and} \bibinfo{person}{Dong Su}.} \bibinfo{year}{2011}\natexlab{}.
\newblock \showarticletitle{Provably private data anonymization: Or,
  k-anonymity meets differential privacy}.
\newblock \bibinfo{journal}{\emph{CoRR, abs/1101.2604}}  \bibinfo{volume}{49}
  (\bibinfo{year}{2011}), \bibinfo{pages}{55}.
\newblock


\bibitem[Lindell and Pinkas(2000)]%
        {lindell2000privacy}
\bibfield{author}{\bibinfo{person}{Yehuda Lindell} {and} \bibinfo{person}{Benny
  Pinkas}.} \bibinfo{year}{2000}\natexlab{}.
\newblock \showarticletitle{Privacy preserving data mining}. In
  \bibinfo{booktitle}{\emph{Annual International Cryptology Conference}}.
  Springer, \bibinfo{pages}{36--54}.
\newblock


\bibitem[Machanavajjhala et~al\mbox{.}(2006)]%
        {1617392}
\bibfield{author}{\bibinfo{person}{A. Machanavajjhala}, \bibinfo{person}{J.
  Gehrke}, \bibinfo{person}{D. Kifer}, {and} \bibinfo{person}{M.
  Venkitasubramaniam}.} \bibinfo{year}{2006}\natexlab{}.
\newblock \showarticletitle{L-diversity: privacy beyond k-anonymity}. In
  \bibinfo{booktitle}{\emph{22nd International Conference on Data Engineering
  (ICDE'06)}}. \bibinfo{pages}{24--24}.
\newblock
\urldef\tempurl%
\url{https://doi.org/10.1109/ICDE.2006.1}
\showDOI{\tempurl}


\bibitem[Marwick and Boyd(2018)]%
        {marwick2018understanding}
\bibfield{author}{\bibinfo{person}{Alice~E Marwick} {and}
  \bibinfo{person}{Danah Boyd}.} \bibinfo{year}{2018}\natexlab{}.
\newblock \showarticletitle{Understanding privacy at the margins.}
\newblock \bibinfo{journal}{\emph{International Journal of Communication
  (19328036)}}  \bibinfo{volume}{12} (\bibinfo{year}{2018}).
\newblock


\bibitem[Nissenbaum(2004)]%
        {nissenbaumCI}
\bibfield{author}{\bibinfo{person}{Helen Nissenbaum}.}
  \bibinfo{year}{2004}\natexlab{}.
\newblock \showarticletitle{Privacy as Contextual Integrity}.
\newblock \bibinfo{journal}{\emph{Washington Law Review}} \bibinfo{volume}{79},
  \bibinfo{number}{1} (\bibinfo{year}{2004}), \bibinfo{pages}{119}.
\newblock


\bibitem[Ohm(2009)]%
        {ohm2009broken}
\bibfield{author}{\bibinfo{person}{Paul Ohm}.} \bibinfo{year}{2009}\natexlab{}.
\newblock \showarticletitle{Broken promises of privacy: Responding to the
  surprising failure of anonymization}.
\newblock \bibinfo{journal}{\emph{UCLA l. Rev.}}  \bibinfo{volume}{57}
  (\bibinfo{year}{2009}), \bibinfo{pages}{1701}.
\newblock


\bibitem[Posner(1977)]%
        {posner1977right}
\bibfield{author}{\bibinfo{person}{Richard~A Posner}.}
  \bibinfo{year}{1977}\natexlab{}.
\newblock \showarticletitle{The right of privacy}.
\newblock \bibinfo{journal}{\emph{Ga. L. Rev.}}  \bibinfo{volume}{12}
  (\bibinfo{year}{1977}), \bibinfo{pages}{393}.
\newblock


\bibitem[Samarati and Sweeney(1998)]%
        {samarati1998protecting}
\bibfield{author}{\bibinfo{person}{Pierangela Samarati} {and}
  \bibinfo{person}{Latanya Sweeney}.} \bibinfo{year}{1998}\natexlab{}.
\newblock \bibinfo{booktitle}{\emph{Protecting privacy when disclosing
  information: k-anonymity and its enforcement through generalization and
  suppression}}.
\newblock \bibinfo{type}{{T}echnical {R}eport}. \bibinfo{institution}{technical
  report, SRI International}.
\newblock


\bibitem[Sandholm(2010)]%
        {sandholm2010decompositions}
\bibfield{author}{\bibinfo{person}{William~H Sandholm}.}
  \bibinfo{year}{2010}\natexlab{}.
\newblock \showarticletitle{Decompositions and potentials for normal form
  games}.
\newblock \bibinfo{journal}{\emph{Games and Economic Behavior}}
  \bibinfo{volume}{70}, \bibinfo{number}{2} (\bibinfo{year}{2010}),
  \bibinfo{pages}{446--456}.
\newblock


\bibitem[Solove(2008)]%
        {soloveUnderstandingPrivacy}
\bibfield{author}{\bibinfo{person}{Daniel~J. Solove}.}
  \bibinfo{year}{2008}\natexlab{}.
\newblock \bibinfo{booktitle}{\emph{Understanding Privacy}}.
\newblock \bibinfo{publisher}{Harvard University Press}.
\newblock


\bibitem[Ulusoy and Yolum(2019)]%
        {ulusoy2019emergent}
\bibfield{author}{\bibinfo{person}{Onuralp Ulusoy} {and}
  \bibinfo{person}{P{\i}nar Yolum}.} \bibinfo{year}{2019}\natexlab{}.
\newblock \showarticletitle{Emergent privacy norms for collaborative systems}.
  In \bibinfo{booktitle}{\emph{PRIMA 2019: Principles and Practice of
  Multi-Agent Systems: 22nd International Conference, Turin, Italy, October
  28--31, 2019, Proceedings 22}}. Springer, \bibinfo{pages}{514--522}.
\newblock


\bibitem[Warner(1965)]%
        {warnerRandomizedResponse}
\bibfield{author}{\bibinfo{person}{Stanley~L. Warner}.}
  \bibinfo{year}{1965}\natexlab{}.
\newblock \showarticletitle{Randomized Response: A Survey Technique for
  Eliminating Evasive Answer Bias}.
\newblock \bibinfo{journal}{\emph{J. Amer. Statist. Assoc.}}
  \bibinfo{volume}{60}, \bibinfo{number}{309} (\bibinfo{year}{1965}),
  \bibinfo{pages}{63--69}.
\newblock
\urldef\tempurl%
\url{https://doi.org/10.1080/01621459.1965.10480775}
\showDOI{\tempurl}


\bibitem[Warren and Brandeis(1890)]%
        {warrenright}
\bibfield{author}{\bibinfo{person}{Samuel~D. Warren} {and}
  \bibinfo{person}{Louis~D. Brandeis}.} \bibinfo{year}{1890}\natexlab{}.
\newblock \showarticletitle{The Right to Privacy}.
\newblock \bibinfo{journal}{\emph{Harvard Law Review}} \bibinfo{volume}{4},
  \bibinfo{number}{5} (\bibinfo{year}{1890}), \bibinfo{pages}{193--220}.
\newblock
\showISSN{0017811X}
\urldef\tempurl%
\url{http://www.jstor.org/stable/1321160}
\showURL{%
\tempurl}


\bibitem[Wuest(2021)]%
        {wuest_2021}
\bibfield{author}{\bibinfo{person}{Joanna Wuest}.}
  \bibinfo{year}{2021}\natexlab{}.
\newblock \showarticletitle{A Conservative Right to Privacy: Legal,
  Ideological, and Coalitional Transformations in US Social Conservatism}.
\newblock \bibinfo{journal}{\emph{Law \& Social Inquiry}} \bibinfo{volume}{46},
  \bibinfo{number}{4} (\bibinfo{year}{2021}), \bibinfo{pages}{964–992}.
\newblock
\urldef\tempurl%
\url{https://doi.org/10.1017/lsi.2021.1}
\showDOI{\tempurl}


\end{thebibliography}

\end{document}